\documentstyle [preprint,aps]{revtex}
\begin{document}
\draft
\title{DISSIPATIVE BOUSSINESQ SYSTEM OF EQUATIONS\\
IN THE B\'ENARD-MARANGONI PHENOMENON}
\author{ R. A. Kraenkel, S. M. Kurcbart, J. G. Pereira}
\address{Instituto de F\'{\i}sica Te\'orica\\
Universidade Estadual Paulista\\
Rua Pamplona 145,\\
01405-900\, S\~ao Paulo\, SP --
Brazil}
\author{M. A. Manna}
\address{Laboratoire de Physique Math\'ematique\\
Universit\'e de Montpellier II\\
34095 Montpellier Cedex 05 --
France}
\date{\today}
\maketitle
\begin{abstract}
By using the long-wave approximation, a system of coupled
evolution equations for the bulk velocity and the surface perturbations
of a B\'enard-Marangoni system is obtained. It includes
nonlinearity, dispersion and dissipation, and it can be interpreted as a
dissipative generalization of the usual Boussinesq system of
equations. As a particular case, a strictly dissipative version of the
Boussinesq system is obtained. Finally, some speculations are made on
the nature of the physical phenomena described by this system of equations.
\end{abstract}

\section{Introduction}

Several recent works \cite{1,2,3,4,5,6,7,8} have dealt with the study of
oscillatory instabilities in systems of the Rayleigh-B\'enard and
B\'enard-Marangoni type. Considering a shallow fluid layer bounded
below by a plane stress-free perfect conducting plate, and with a
deformable upper free surface where a constant heat flux is
imposed, the linear stability analysis indicates that oscillatory
motion sets in when a certain critical combination of the Rayleigh
and Marangoni numbers is attained~\cite{1}. Subsequently, it was
shown that, when the surface-tension can be disregarded and the
Rayleigh number of the system is at $R = 30$, appropriate surface
deflections governed by the Korteweg-de Vries (KdV)
equation appear~\cite{2}. For a two-dimensional surface, the same
deflections turn out to be governed by the Kadomtsev-Petviashvili
(KP) equation~\cite{3}. Thereafter, these results have been extended
to the B\'enard-Marangoni system~\cite{4,5}, where buoyancy is
discarded. In this case, appropriate surface disturbances propagate
according to the KdV equation when the Marangoni number of the
system is at $M = -12$. Further extentions to a double-diffusive
systems~\cite{6}, and to a temperature depending
viscosity~\cite{7} have also been established. The physical
mechanism behind such sustained excitations is the balance, at the
critical point, between the energy dissipated by viscous forces and
that released either by buoyancy, or by a temperature depending
surface tension. In both cases, out of the critical points,
appropriate surface excitations are governed by the Burgers
equation~\cite{5,8}.

The KdV equation is well known to govern long surface-waves in
inviscid shallow fluids~\cite{9}. It corresponds to a situation
where nonlinearity and dispersion compensate each other, making
possible the existence of coherent wave-structures, like the
solitary-wave. The reductive perturbation method of Taniuti~\cite{10},
based on the concept of stretching, and in which waves propagating
in only one direction are sought from the beginning, is a common
approach to study long waves in shallow water. From the various
possible stretchings of the coordinates, different evolution
equations  (linear wave, breaking wave, KdV) may emerge as governing
surface disturbances. However, there is an alternative approach to
the theory of long waves in shallow water, which is based on
perturbative expansions in two small parameters~\cite{9}. One of
them measures the smallness of the amplitude perturbation, and the
other is a measure of the longness of the wave-length perturbation.
This approach, as an intermediate step, and from different possible
relations between the two perturbative parameters, yields
different systems of evolution equations (nonlinear shallow water,
Boussinesq) describing superimposed waves propagating in
opposite directions. Only when specializing to waves moving in a
given direction, equations like breaking wave and KdV show up. A
perturbative scheme of this kind has not been used to study
surface excitations in B\'enard systems. Consequently, for these
systems,  the more general evolution equations, wherefrom Burgers
and KdV equations are obtained, have not been found.

In this paper, instead of using the reductive perturbation method
of Taniuti~\cite{10}, we will proceed through a perturbation
scheme for the B\'enard-Marangoni system based on
two perturbative parameters, leading to the so called long-waves in
shallow water approximation. By this way, a new system of coupled
evolution equations will be found, involving the fluid velocity and
the free-surface displacement. This system may be interpreted as a
dissipative generalization of the Boussinesq equations. When the
Marangoni number assumes the critical value $M = -12$, and a certain
relation between the perturbative parameters is assumed, it reduces,
as we are going to see, to the usual Boussinesq equations. A further
restriction to waves moving in only one direction leads to the KdV
equation. On the other hand, out of the critical point, and assuming
a different relation between the perturbative parameters, the
dissipative generalization of the Boussinesq equations reduce to a
strictly dissipative version of the Boussinesq equations. In this
case, a restriction to waves moving in only one direction will lead
to the Burgers equation. Although we have restricted ourselves to
the B\'enard-Marangoni system, an extension to the Rayleigh-B\'enard
case, where buoyancy is predominant, may also be established with
similar results~\cite{11}. Finally, through a linear analysis, a
speculation on the nature of the physical solutions to the strictly
dissipative Boussinesq system of equations is made.

\section{Basic Equations and Boundary Conditions}

We consider a fluid bounded below by a plane, stress-free, perfect
termally conducting plate at $z = 0$, and above by a deformable
one-dimensional free surface which, at rest, lies at $z = d$. The
depth $d$ will be supposed to be small enough so that buoyancy can
be neglected when compared to the effects coming from the surface
tension dependence on temperature. In other words, we will be
dealing with a B\'enard-Marangoni system. The equations that
describe such a system are
\begin{equation}
 {\vec \nabla} . {\vec v} = 0 \, ,
\end{equation}
\begin{equation}
 {\rho {d\vec v \over dt}} = -{\vec {\nabla}p + \mu \nabla^2 \vec v
+ \vec g \rho} \, ,
\end{equation}
\begin{equation}
 {dT \over dt} = {\kappa \nabla^2 T} \, ,
\end{equation}
where ${d \over dt} = {\partial \over \partial t} + \vec v
. \vec {\nabla}$ is the convective derivative, $\vec v = (u, 0, w)$
is the fluid velocity, and $p$ is the pressure. The density
$\rho$, the viscosity $\mu$, and the thermal diffusivity coefficient
$\kappa$ are supposed to be constant. The surface tension $\tau$
will be assumed to depend linearly on the temperature:
\begin{equation}
 \tau = {\tau_0 [1 - \gamma (T - T_0)]} \, ,
\end{equation}
where $\gamma$ is a constant, and $\tau_0$, $T_0$ are reference
values for the surface tension and the temperature, respectively.

The boundary conditions on the upper free surface $z = d + \eta
(x,t)$ are~\cite{12}
\begin{equation}
 \eta_t + u \eta_x = w \, ,
\end{equation}
\begin{equation}
 \left(p - p_a \right) - {2 \mu \over N^2} \left(w_z + u_x
 \eta_x^2 - \eta_x u_z - \eta_x w_x \right) = - {\tau \over N^3}
 \eta_{xx} \, ,
\end{equation}
\begin{equation}
 \mu\left(1 - \eta_x^2 \right) \left(u_z + w_x \right) + 2 \mu
 \eta_x \left(w_z - u_x \right) = N \left(\tau_x + \eta_x \tau_z
 \right) \, ,
\end{equation}
\begin{equation}
 \eta_x T_x - T_z = {F \over k} N \, ,
\end{equation}
where $F$ is the normal heat flux, $k$ is the thermal conductivity,
$p_a$ is a pressure exerted on the upper surface, all of them
supposed to be constant, and $N = \left(1 + \eta_x^2 \right)^{1
\over 2}$. Subscripts denote partial derivatives with respect to the
corresponding coordinate.

On the lower plane $z = 0$, we assume that the sliding resistance
between two portions of the fluid is much greater than between the
fluid and the plane~\cite{13}, implying a stress-free lower surface:
\begin{equation}
 w = u_z = 0 \, .
\end{equation}
Moreover, we will assume that the lower plate is at a constant
temperature $T = T_b$.

\section{Perturbative Solution: Evolution Equations}

The static solution to the above equations is given by
$$p_s = p_a - \rho g (z - d) \, ,$$
$$T_s = T_0 - {F \over k} (z - d) \, .$$
We consider now perturbations from this quiescent state. The
horizontal and vertical length scales of these perturbations are
supposed to be $l$ and $a$, respectively. Then, we define two small
parameters
\begin{equation}
 \epsilon = {a \over d} \hskip 3truecm \delta = {d \over l} \, ,
\end{equation}
which will be used to order the expansions. Before proceeding
further, however, it is convenient to write the equations,
boundary conditions and static solutions in a dimensionless form.
This is done by taking the original variables (primed) to be
\begin{mathletters}
\begin{eqnarray}
 x^\prime &=& l x \hskip 1truecm z^\prime = d z \hskip
 1truecm t^\prime = {l\over c_0} \, t \\
 \eta^\prime &=& a \eta \hskip 1truecm u^\prime = {a g \over
 c_0} \, u \hskip 1truecm
 w^\prime = {a g \over c_0 \delta} \, w \, ,
\end{eqnarray}
\end{mathletters}
where $c_0^2 = g d$. Furthermore, four dimensionless parameters
appear: the Pradtl number $\sigma = \mu / \rho \kappa$; the
Reinolds number $R = c_0 d \rho / \mu$; the Bond number $B =
\rho g d^2 / \tau_0$, and the Marangoni number $M = \gamma F
d^2 \tau_0 / k \kappa \mu$.

To obtain the nonlinear evolution of the surface perturbations in
the shallow water theory, we expand all variables in powers of
$z$, keeping the terms that will contribute to the evolution
equations up to orders $\epsilon$ and $\delta^2$. Despite laborious,
this procedure is straightforward, and for this reason we will omit
the details, specifying in the text only the general guidelines, and
giving a brief summary of the results in the Appendix. To start
with, we make the expansions
$$u = \sum _{n=0}^\infty z^n u_n \hskip 1truecm w =
\sum_{n=0}^\infty z^n w_n \, ,$$
where $u_n$ and $w_n$ are both functions of $x$ and $t$. Then,
substituting them in Eq.(1), we get the relation
$$w_{n+1} = {- \delta^2 {u_{nx} \over n+1}} \, .$$
Using the boundary conditions at $z = 0$, it is easy to show that
$$w_0 = w_2 = w_4 = ... = 0 \,,$$
$$u_1 = u_3 = u_5 = ... = 0 \, .$$
Then, using the expansion
$$p = p_s + \sum _{n=0}^\infty z^n p_n$$
in Eq. (2), it is possible to obtain $u_2$, $u_4$, $u_6$ and
$p_2$ in terms of $u_0$ and $p_0$ only. The other components of
the expansions will contribute to orders higher than $\epsilon$ and
$\delta^2$, and therefore they can be neglected.

Next, expanding the temperature according to
$$T = T_s + \sum _{n=0}^\infty z^n \theta_n \, ,$$
and using Eq. (3) with the corresponding boundary conditions, we
can see that
$$\theta_0 = \theta_2 = \theta_4 = \theta_6 = ... = 0 \, .$$
Moreover, we can also obtain expressions for $\theta_3$ and
$\theta_5$ in terms of $u_0$ and $p_0$ only. Now, Eq.(6) yields
$p_0$ in terms of $u_0$ and $\eta$. Consequently, it is possible
to rewrite the $u$'s, $p$'s and $\theta$'s in terms of
$u_0$ and $\eta$ only. Finally, using the above results in Eqs. (5)
and (7), we obtain, up to order $\epsilon$ and $\delta^2$, a
coupled system of evolution equations for $u_0$ and $\eta$:
\begin{eqnarray}
 u_{0t} &+& \epsilon u_0 u_{0x} + c^2 \eta_x - {\delta \over R}
 \left(4 + {M \over 3} \right) u_{0xx} + {\delta R \over 6}
 \left(u_{0tt} + \eta_{xt} \right) + {\delta^2 R^2 \over120}
 \left(u_{0ttt} + \eta_{xtt} \right) \nonumber \\
 &-& \delta^2 \left[{11 \over 6} -
 {M \over 30} \left(4 \sigma - 1 \right) \right] u_{0xxt} -
 \delta^2 \left({1 \over B} + {M \over
 30} + 1 \right) \eta_{xxx} = 0 \, ,
\end{eqnarray}
\begin{eqnarray}
 \eta_t + u_{0x} + \epsilon \left(u_0 \eta \right)_x + {\delta R
 \over 6} \left(u_{0xt} + \eta_{xx} \right) - {\delta^2 \over 2}
 u_{0xxx} + {\delta^2 R^2 \over 120} \left(u_{0xtt} + \eta_{xxt}
 \right) = 0 \, ,
\end{eqnarray}
where
\begin{equation}
 c^2 = 1 - {M \over \sigma R^2} \, .
\end{equation}

The velocity $u_0$ is only the first term in the expansion of
$u$, which is :
$$u = u_0 + \delta {z^2 R^2 \over 2} \left(u_{0t} + \eta_x - {3
\delta \over R}u_{0xx} \right) + \delta^2 {z^4 R^2 \over 24}
\left(u_{0tt} + \eta_{xt} \right) + {\cal O} \left(\epsilon \delta ,
\delta^3 \right) \, .$$
The value averaged over the depth is
$${\tilde u} = u_0 + \delta {R \over 6} \left(u_{0t} + \eta_x
\right) - {\delta^2 \over 2} \left[u_{0xx} - {R^2 \over 60}
\left(u_{0tt} + \eta_{xt} \right) \right] + {\cal O} \left(\epsilon
\delta , \delta^3 \right) .$$
The inverse is
$$u_0 = \tilde u - \delta {R \over 6} \left(\eta_x + \tilde u_t
\right) + {\delta^2 \over 2} \left[\tilde u_{xx} + {7 R^2 \over 180}
\left( \eta_{xt} + \tilde u_{tt} \right) \right] + {\cal O}
\left(\epsilon \delta , \delta^3 \right) \, .$$
Substituting this in Eqs. (12) and (13), and using the lowest
order equations into the $\tilde u_{xxt}$ term, we obtain (omitting
the tilde):
\begin{mathletters}
\begin{eqnarray}
 u_t &+& c^2 \eta_x + \epsilon u u_x - {\delta \over R} \left(4
 + {M \over 3} \right) u_{xx} + \delta^2 \Lambda \eta_{xtt} = 0 \\
 \eta_t &+& u_x + \epsilon (u \eta)_x = 0 \, ,
\end{eqnarray}
\end{mathletters}
where
\begin{equation}
 \Lambda = {4 \over 3} - {1 \over c^2} \left(1 + {1 \over B}
 \right) + {M \over 30} \left(1 - {1 \over c^2} - 4 \sigma
 \right) + {1 \over 6} \left(4 + {M \over 3} \right) \left(1 -
 {1 \over c^2} \right) .
\end{equation}
Due to the presence of the $u_{xx}$ term in Eqs. (15), this
system can be considered as a dissipative generalization of
the Boussinesq equations. When the Marangoni number is at
the critical value $M = - 12$, these equations coincide with the
usual Boussinesq system of equations
\begin{mathletters}
\begin{eqnarray}
 u_t &+& c^2 \eta_x + \epsilon u u_x + \delta^2 \Lambda \eta_{xtt} =
 0 \\
 \eta_t &+& u_x + \epsilon (u \eta)_x = 0 \, .
\end{eqnarray}
\end{mathletters}
In this case, by assuming
that $\delta^2 \approx \epsilon$, and by specializing to
waves moving, say, to the right, we can obtain a relation between
$u$ and $\eta$:
\begin{equation}
 u = c \eta - \epsilon {c \over 4} \, \eta^2 - \delta^2 {\Lambda
 \over 2} \, \eta_{xt} \, ,
\end{equation}
which is a kind of Riemann invariant~\cite{9}. Using this in Eq.
(15) yields
\begin{equation}
 \eta_t + c \eta_x + \epsilon {3 c \over 2} \eta \eta_x +
 \delta^2 {\Lambda \over 2} \eta_{xxx} = 0 \, ,
\end{equation}
where now
\begin{equation}
 \Lambda = {1 \over5} \left({14 \over 3} + 8 \sigma \right) -
 {1 \over c^2} \left({3 \over 5} + {1 \over B} \right) \, .
\end{equation}
Equation(19) is the KdV equation.

Let us now consider the case $M \not = - 12$. Assuming that
$\delta \approx \epsilon$, and neglecting terms of order
$\delta^2 \approx \epsilon^2$, Eq. (15) becomes:
\begin{mathletters}
\begin{eqnarray}
 u_t &+& c^2 \eta_x + \epsilon u u_x - {\delta \over R} \left(4
 + {M \over 3} \right) u_{xx} = 0 \\
 \eta_t &+& u_x + \epsilon (u \eta)_x = 0 \, .
\end{eqnarray}
\end{mathletters}
This is a strictly dissipative version of the Boussinesq
system, since the dispersive term is not present now. By
specializing again to waves moving to the right, we obtain
the following relation between $u$ and $\eta$:
\begin{equation}
 u = c \eta - {\epsilon c \over 4} \, \eta^2 - {\delta \over 2 R}
 \left(4 + {M \over 3} \right) \eta_x \, .
\end{equation}
Substituting this relation in Eq. (21), we obtain
\begin{equation}
 \eta_t + c \eta_x + \epsilon {3 c \over 2} \eta
 \eta_x - {\delta \over 2 R} \left(4 + {M \over 3} \right) \eta_{xx}
 = 0 \,,
\end{equation}
which is the Burgers equation.
\section{THE STRICTLY DISSIPATIVE BOUSSINESQ EQUATION}

The usual Boussinesq system of equations, Eqs. (17), may be transformed
into the $classical$ Boussinesq equations, also known as
dispersive long-wave equations ~\cite{14}. These equations have already been
shown to be integrable~\cite {15}. The solutions to KdV and
Burgers equations have also been extensively
discussed in the literature~\cite {16}. However, the strictly dissipative
Boussinesq system of equations, Eqs. (21), seems not to have been handled. The
existence of analytical solutions, therefore, is still an open issue.

The purpose of this section is to make some speculations on the nature of the
physical phenomena governed by the strictly dissipative Boussinesq system of
equations. To start with, we first rewrite Eqs. (21) in a dimensional form:
\begin{mathletters}
\begin{eqnarray}
u_t &+& c^2 g \eta_x + uu_x - \alpha d c_0 u_{xx} = 0 \\
\eta_t &+& d u_x + u_x \eta + u\eta_x = 0 \,,
\end{eqnarray}
\end{mathletters}
with
$$\alpha = {1 \over R} \left(4+{M \over3} \right) \, ,$$
and $c^2$ given by Eq. (14). For $M=-12$, and changing the origin of the
coordinates according to
$$\eta(x,t) = h(x,t) - d \, ,$$
we obtain
\begin{mathletters}
\begin{eqnarray}
u_t + c^2 g h_x + u u_x = 0 \\
h_t + u_x h + u h_x = 0
\end{eqnarray}
\end{mathletters}
which, when $c^2=1$, coincides with the classical
one-dimensional shallow water system of equations~\cite{9}. A general
analytical solution to this system, which might present shocks, has
already been reported in the
literature~\cite {17}. In fact, by restricting to
waves moving to the right we obtain the breaking-wave equation~\cite {9}
\begin{equation}
h_t - {c \over2} \sqrt{g d}\, h_x + {3 c \over2} \sqrt{g \over d}\,
h h_x = 0 \,.
\end{equation}

On the other hand, for $M \not = - 12$, it is possible to
gain some insight on the nature of the solutions to
Eqs. (24) by making a linear analysis. Neglecting the nonlinear
term, Eqs. (24) reads:
\begin{mathletters}
\begin{eqnarray}
u_t &+& c^2 g \eta_x - \alpha d c_0 u_{xx} = 0 \\
\eta_t &+& d u_x = 0 \, .
\end{eqnarray}
\end{mathletters}
Supposing solutions of the form $exp(\kappa x + \omega t)$, we obtain the
following dispersion relation:
\begin{equation}
\omega = i{{\alpha d c_0 \kappa^2} \over2} \pm {{\alpha d c_0 \kappa^2}\over2}
\left[ -1 + {4 c^2\over \alpha^2 d^2 \kappa^2} \right]^{1/2} \, .
\end{equation}
There are three cases to be considered: \\
(i)When ${4 c^2}/{\alpha^2 d^2 \kappa^2} < 1$, then
$$\kappa > {2c\over\alpha d}\qquad \mbox{or} \qquad \kappa
< - { 2c\over\alpha d}\,,$$
and the frequency $\omega$ is a pure imaginary number.
In this case the system will present either, a strictly  dissipative or
antidissipative behavior, depending on whether the imaginary part
of $\omega$ is negative or positive.\\
(ii)When ${4c^2}/{\alpha^2 d^2 \kappa^2} >1$, then
$$- {2c \over\alpha d} < \kappa < {2c \over\alpha d} \, ,$$
and the frequency $\omega$ is a complex number. In this case the system will
present oscillatory motion combined with either dissipation or
anti-dissipation, depending on whether the imaginary part of $\omega$ is
negative or positive. For $\kappa$ within this interval, the
nonlinear dissipative
Boussinesq system, Eqs. (24), may possibly present a constant-profile opposite
two travelling-wave solutions if a compensation mechanism between dissipation
and nonlinearity takes place, in a way similar to what happens to the Taylor
shock profile solution of Burgers equation~\cite {9}. This, however, is
still an open question.\\
(iii)When ${4 c^2}/{\alpha^2 d^2 \kappa^2} = 1$, then
$$\kappa = \pm {2c \over{\alpha d}} \, ,$$
and the frequency is again a pure imaginary number, $\omega= i \omega_I$, with
$$\omega_I = {1\over2} \alpha d c_0 \kappa^2 \, . $$
This is a transition case between the two cases previously described, in which
the system will again present either a strictly dissipative or anti-dissipative
behavior, depending on whether $\omega_I$ is negative or positive. For these
values of $\kappa$, as well as for the values of case (i), the nonlinear
dissipative Boussinesq system will present no oscillatory motion, which means
that its solutions must describe pure nonlinear diffusive phenomena at this
region.
\section{FINAL COMMENTS}

The results obtained in this paper are two-fold. First, from a perturbative
analysis, corresponding to the long-wave in shallow-water approximation, we
obtained a dissipative generalization of the Boussinesq system of equations as
governing the bulk velocity and surface perturbations of a
B\'enard-Marangoni phenomenon. This system of coupled evolution equations
includes nonlinearity, dispersion and dissipation. The predominance
of any one of them depends on the relation between $\epsilon$ and
$\delta$. Three cases are of special interest. First, when $M = -
12$, the dissipative term vanishes, and assuming that $\delta^2
\approx \epsilon$, the usual Boussinesq system of equations is
obtained. Second, when the Marangoni number is far enough from the
critical value, that is when $M + 12 = {\cal O} (1)$, and assuming
now that $\delta \approx \epsilon$, the dissipative term dominates
over the dispersive one. Then, by neglecting the dispersive term we
got Eq. (21), which is a strictly dissipative version of the
Boussinesq system of equations. And finally, as an intermediate case, we
may also consider the situation in which $M + 12 = {\cal O}
(\delta)$. Assuming again that  $\delta^2 \approx \epsilon$, it
results that the dispersive and dissipative terms of Eq. (15) are
of the same order. In this case, no term is neglected, and the
generalized Boussinesq system, Eq. (15), will govern the bulk velocity
and surface perturbations of the B\'enard-Marangoni system. Upon
specialization to waves moving, say, to the right, each one of these
three cases will lead to a single evolution equation for the surface
displacement, which will be, respectively, the KdV, Burgers and
KdV-Burgers equations. It should be remarked that the main feature
of the perturbative
scheme adopted here was to allow for a clear distinction of the
relative importance of each term in Eq.(15). By considering different
relations between the parameters $\delta$ and $\epsilon$, a
criteria to collect terms of the dominant order, and to neglect
higher order ones, was thus established.

The second result of this paper refers to the appearance of a new equation,
which we have called the strictly dissipative Boussinesq system of equations.
It
seems to be a nonintegrable equation, and the existence of analytical solutions
is still an open question. Despite of this problem, we have performed here
a linear analysis to get some insight on the nature of the physical
phenomena it might describe.
Finally, due to the fact that it shows up in a physical system, we
believe this system of equation deserves further investigations as only few
results exist with respect to dissipative systems.

\acknowledgements

The authors would like to thank Conselho Nacional de
Desenvolvimento Cient\'{\i}fico e Tecnol\'ogico (CNPq), Brazil, for
partial support. One of the authors (MAM) would like also to thank
the Instituto de F\'{\i}sica Te\'orica, UNESP, for the kind
hospitality, and the CNPq for a travel grant.

\appendix
\section*{}

We give here a summary of the results described, but not
explicitly shown in the text. The expressions for each term of the
velocity expansion, already written in terms of $u_0$ and $p_0$,
and considering only those terms that will contribute to the
evolution equations up to orders $\epsilon$ and $\delta^2$, are
$$u_2  = {\delta R \over 2} \left(u_{0t} + \epsilon u_0
u_{0x} + {1 \over \epsilon} p_{0x} - {\delta \over R} u_{0xx}
\right) \, ,$$
$$u_4  = {\delta^3 R \over 24} \left( - {2 \over \epsilon} p_{0xxx}
+ {R \over \delta} u_{0tt} + {R \over \epsilon \delta} p_{0xt} - 2
u_{0xxt} \right) \, ,$$
$$u_6  = {\delta^3 R^3 \over 720} \left(u_{0ttt} + {1 \over
\epsilon} p_{0xtt} \right) \, . $$
The next terms of the expansion will not contribute to the
evolution equations. In the same way, the only term of the
pressure expansion that will contribute is
$$p_2 = - {\delta^2 \over 2} p_{0xx} \, .$$
{}From Eq. (3), with the corresponding boundary conditions, we can
obtain expressions for $\theta_1$, $\theta_3$ and $\theta_5$ in
terms of $u_0$ and $p_0$ only:
$$\theta_1  = - \epsilon \delta {R \sigma \over 2} u_{0x}
+ \epsilon \delta^2 {R^2 \sigma \over 24} (5 \sigma - 1) u_{0xt} -
\delta^2 {R^2 \sigma \over 24} p_{0xx} \, ,$$
$$\theta_3  = - \epsilon \delta^2 {R^2 \sigma^2 \over 12} u_{0xt} +
\epsilon \delta^3 {R^2 \sigma^2 \over 24} (5 \sigma - 1) u_{0xtt}
+ \epsilon \delta {R \sigma \over 6} u_{0x}\, , $$
$$\theta_5  = \epsilon \delta^2 \left[ \left(1 + {1 \over \sigma}
\right) u_{0xt} + {1 \over \epsilon \sigma} p_{0xx} - \delta {R
\sigma \over 2} u_{0xtt} - \delta {2 \over R \sigma} u_{0xxx}
\right] \, .$$
Again, these are the only components of the temperature expansion
that will contribute, up to orders $\epsilon$ and $\delta^2$, to the
evolution equations.

Now, from Eq. (6), we can get $p_0$ in terms of $u_0$ and $\eta$
only:
$$p_0 = \epsilon \eta - \epsilon \delta {2 \over R} u_{0x} -
\epsilon \delta^2 \left[ u_{0xt} + \left({1 \over B}  + {1 \over
2} \right) \eta_{xx} \right] + \epsilon \delta^3 \left[ {2 \over
R} u_{0xxx} - {R \over 12} u_{0xtt} \right] \, .$$
This equation allows us to rewrite the $u$'s, $p$'s and $\theta$'s
in terms of $u_0$ and $\eta$. Consequently, the system of
evolution equations, Eq. (12), obtained from Eqs. (5) and (7)
could also be written in terms of $u_0$ and $\eta$ only.

\end{document}